\documentclass{pasj00}
\draft

\begin{document}
\SetRunningHead{Miyata et al.}{Resonance Line Scattering in Cygnus Loop}
\Received{2007/06/31}%{yyyy/mm/dd}
\Accepted{2007/10/08}%{yyyy/mm/dd}

\title{Evidence for Resonance Line Scattering in the Suzaku X-ray
Spectrum of the Cygnus Loop}

%%% begin:list of authors
\author{
Emi \textsc{Miyata}\altaffilmark{1},
Kuniaki \textsc{Masai}\altaffilmark{2}
\and
John P. \textsc{Hughes}\altaffilmark{3}

\altaffiltext{1}{Osaka University}
\email{miyata@ess.sci.osaka-u.ac.jp}
\altaffiltext{2}{Tokyo Metropolitan University}
\email{masai@phys.metro-u.ac.jp}
\altaffiltext{3}{Rutgers University}
\email{jph@physics.rutgers.edu}
}
%%% end:list of authors

%%% Please use the following style in case that sorting by 
%%% affiliation is impossible. 
%
% \author{%
%   D-Firstname \textsc{D-Familyname}\altaffilmark{1}
%   E-Firstname \textsc{E-Familyname}\altaffilmark{1,2}
%   and
%   F-Firstname \textsc{F-Familyname}\altaffilmark{2}}
% \altaffiltext{1}{Address of Institute}
% \email{ddddd@xxx.xxx.xx.xx}
% \email{eeeee@xxx.xxx.xx.xx}
% \altaffiltext{2}{Address of Institute}

%% `\KeyWords{}' always has to be placed before `\maketitle'.
 %Do NOT move this preamble from here!

\maketitle

\begin{abstract}
 We present an analysis of the Suzaku observation of the northeastern
 rim of the Cygnus Loop supernova remnant.  The high detection
 efficiency together with the high spectral resolution of the Suzaku
 X-ray CCD camera enables us to detect highly-ionized C and N emission
 lines from the Cygnus Loop.  Given the significant plasma structure
 within the Suzaku field of view, we selected the softest region based
 on ROSAT observations.  The Suzaku spectral data are well
 characterized by a two-component non-equilibrium ionization model with
 different best-fit values for both the electron temperature and ionization
 timescale.  Abundances of C to Fe are all depleted to typically 0.23
 times solar with the exception of O.  The abundance of O is relatively
 depleted by an additional factor of two compared with other heavy
 elements.  We found that the resonance-line-scattering optical depth
 for the intense resonance lines of O is significant and, whereas the
 optical depth for other resonance lines is not as significant, it still
 needs to be taken into account for accurate abundance determination.
\end{abstract}

 \section{Introduction}

 X-ray spectroscopic imaging observations of supernova remnants (SNRs)
 enable us to investigate the thermodynamic state of the hot plasma 
 they contain.  The spatial
 structure of electron temperature ($kT_{\rm e}$), ionization timescale,
 and elemental abundance have now been derived for many SNRs with
 ASCA, Chandra, XMM-Newton, and Suzaku.  The X-ray emission from SNRs is
 usually assumed to be optically thin since the electron density is typically 
 low (0.1$-$10\,cm$^{-3}$).  This is safely the case for continuum
 emission, however the optical depth for the resonance lines cannot be
 assumed to be negligibly small for bright SNRs, as first pointed
 out by Kaastra \& Mewe (1995) in the context of  X-ray observations.
 Resonance transitions are optically allowed
 transitions from the ground state of an ion.  In the coronal limit, they 
 are more likely to occur
 than other transitions, as there is a large population of ions in the
 ground state. If there is a sufficient ion column density along a 
 particular line
 of sight, resonance line photons can be scattered out of that line of sight
 to appear at another location. Resonance
 line scattering, thus, leads to an underestimate of elemental
 abundances as well as biases in $kT_{\rm e}$ determined by 
 line intensity ratios.
 It should be noted that any random or turbulent velocities will tend to
 decrease the effects of resonance line scattering.

 The Cygnus Loop is one of the brightest SNRs in the soft X-ray band
 and appears as a rim-brightened shell.  The Cygnus Loop is nearby
 (540\,pc; Blair et al.~2005) and has a low neutral H column density
 (several $\times 10^{20}$\,cm$^{-2}$) and large apparent size
 (2.$\!\!^\circ$5$\times$3.$\!\!^\circ$5; Levenson et al.~1997), which
 enable us to study its spatially-resolved soft X-ray emission.
 Analysis of IUE observations of the Cygnus Loop
 indicates a high shock velocity (130\,km s$^{-1}$), departures from
 steady flow behind the shock, and significant optical depths for the
 UV resonance lines (Raymond et al.~1981).  Subsequent studies (e.g.,
 Raymond et al.~1988; Cornett et al.~1992) have shown that optical
 depth effects play an important role in shaping the UV and optical spectral
 and morphological properties of the Cygnus Loop.

 In this paper, we report on the very soft X-ray emission of the
 northeastern  region of the Cygnus Loop as observed with the Suzaku
 observatory (Mitsuda et al.~2007).  The extended low energy
 range of the Suzaku X-ray CCD camera (XIS; Koyama et al.~2007)
 combined with their superior energy resolution allows us to detect
 highly ionized C and N emission lines from the northeastern region of
 the Cygnus Loop (Miyata et al.~2007;
 paper I, hereafter).  These authors determined the abundances of C, N, and O
 independently for the first time for this source and found that their
 relative abundances are roughly consistent with solar values.
 Throughout the present paper, CGS units are used unless specified otherwise.

 \section{Observations and Data Screening}

 We observed the northeastern region of the Cygnus Loop supernova remnant
 with the Suzaku observatory on Nov.~24, 2005, during the time allocated
 to the science working group (observation ID
 500021010).  In this paper, we present results
 obtained with the back-illuminated CCD in order to utilize
 its substantially improved sensitivity at energies below 2\,keV compared
 to the front-illuminated CCDs.
 Although the data employed in this paper are the same as those in 
 paper I, here we apply the latest
 calibration files (revision 0.7), which allow, most importantly, for the 
 correction of charge-transfer
 inefficiency (CTI) effects. For this we employed the CTI correction file
 {\tt ae\_xi1\_makepi\_20060522.fits}.  We subtracted a blank-sky spectrum
 obtained from the Lockman hole (observation ID 100046010) since its
 observation date (Nov.~14, 2005) was close to that
 of the Cygnus Loop.
 
 We used {\tt xisrmfgen} version 2006-10-26 to make the spectral response
 matrix file (RMF)  (Ishisaki et al.~2007), including
 the time-dependent degradation of the XIS energy
 resolution (which was  not taken into account in paper I).
 We employed the Monte Carlo
 simulation software {\tt xissimarfgen} version 2006-10-26 to calculate
 the ancillary response file (ARF), taking into account the
 degradation of detection efficiency caused by the build-up of
 contamination on the front of the XIS cameras (Koyama et al.~2007).
 We ignored the energy range of 1.7--1.9\,keV since the gain calibration
 near the Si-edge of the detector has not yet been accurately calibrated.

 \section{Results}

 Spatially-resolved analysis with ASCA revealed significant variation
 of the plasma structure at the northeastern region of the Cygnus
 Loop (Miyata et al.~1994).  Paper I presented results from spectral 
 analysis of seven annular regions spanning the northeastern rim.  
 We focus here on the emission region
 with the lowest electron temperature ($kT_{\rm e}$), for which the emission 
 predominantly comes from highly ionized atoms of
 C, N, and O.  Since the ROSAT PSPC had large effective
 area below the C edge with moderate energy resolution as well as
 good spatial resolution, we selected the region for further Suzaku analysis
 based on the PSPC data. 

\begin{figure}[htbp]
  \centering
  \includegraphics[clip,scale=0.5]{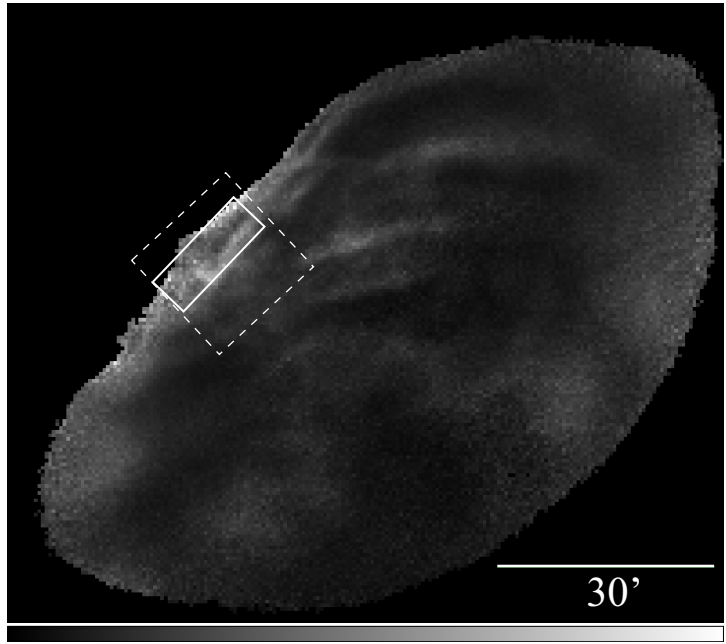}
  \caption{Band-ratio image of 1/4\,keV band to 3/4\,keV band
  at the northeastern region of the Cygnus Loop obtained with the
  ROSAT PSPC. White color represents enhanced 1/4\,keV
  band emission.  The dotted square shows the XIS field of view.
  The region extracted spectrum is shown by the white
  rectangular.}
  \label{fig:image}
 \end{figure}

 We retrieved the PSPC data (observation ID  500034) from the High Energy
 Astrophysics Science Archive  Research Center Online Service, provided
 by the NASA/Goddard Space  Flight Center.  
 Figure~\ref{fig:image} shows the image made by taking the ratio of flux 
 in the  1/4\,keV and  3/4\,keV bands.  
 As clearly shown in this image, there is
 significant spectral variation over the ROSAT field of view (FOV) in both
 the radial and azimuthal directions, with the 
 very softest X-ray emission concentrated at the shock front.  We thus
 extracted the Suzaku XIS spectrum from the region with the softest apparent
 X-ray emission (indicated by
 the white rectangular in figure~\ref{fig:image}).
 The size of this extraction region is
 16.$\!^\prime$2$\times$5.$\!^\prime$9.

\begin{figure}[htbp]
  \centering
  \rotatebox{-90}{\includegraphics[clip,scale=0.5]{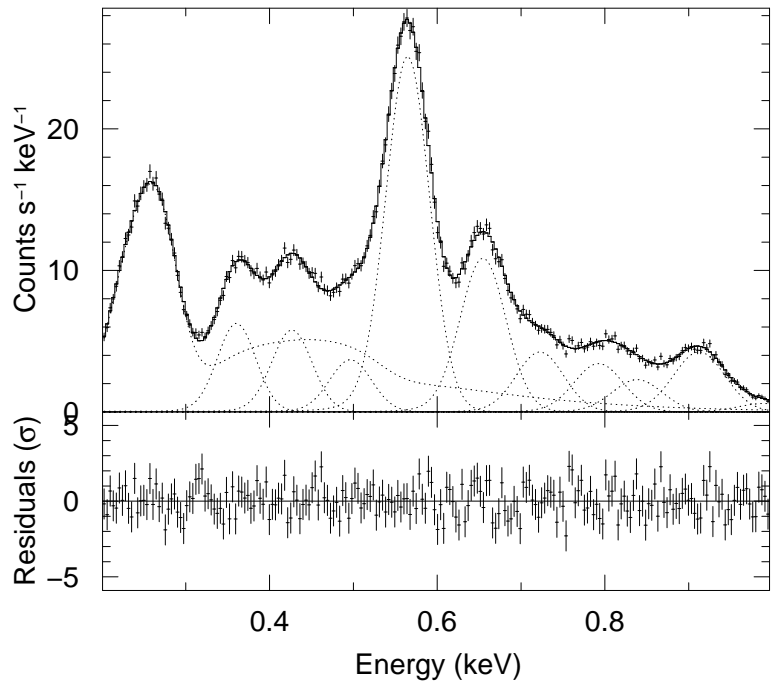}}
  \caption{Spectrum extracted from the softest region shown in
  figure~\ref{fig:image}. The best-fit curve of 10 Gaussian functions with
  the thermal bremsstrahlung model is shown by solid line.  The
  contribution of each component is also shown by dotted line.
  Residuals between data and the best-fit model are shown in the lower
  panel.}
  \label{fig:spec}
 \end{figure}

\begin{figure}[htbp]
  \centering
  \rotatebox{-90}{\includegraphics[clip,scale=0.5]{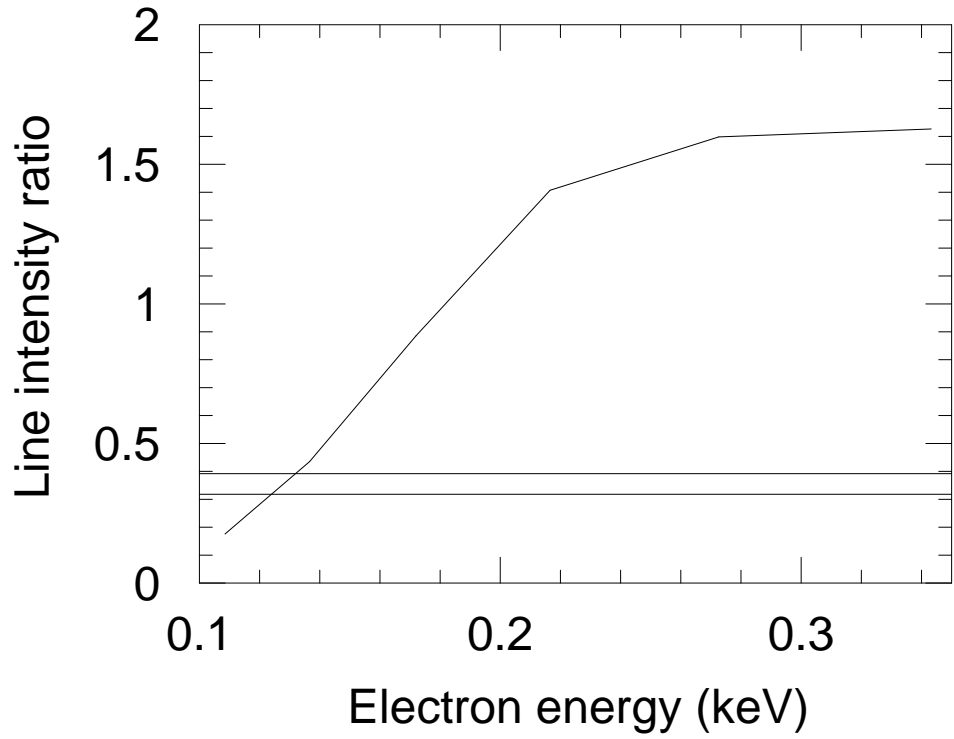}}
  \caption{Line intensity ratio of 0.48$-$0.51\,keV band to
  0.41$-$0.44\,keV band. Two horizontal lines show the upper and lower
  limits for the intensity ratio of 0.495\,eV line to 0.426\,eV line.}
  \label{fig:line_intensity_ratio:N}
 \end{figure}

\begin{figure}[htbp]
  \centering
  \rotatebox{-90}{\includegraphics[clip,scale=0.5]{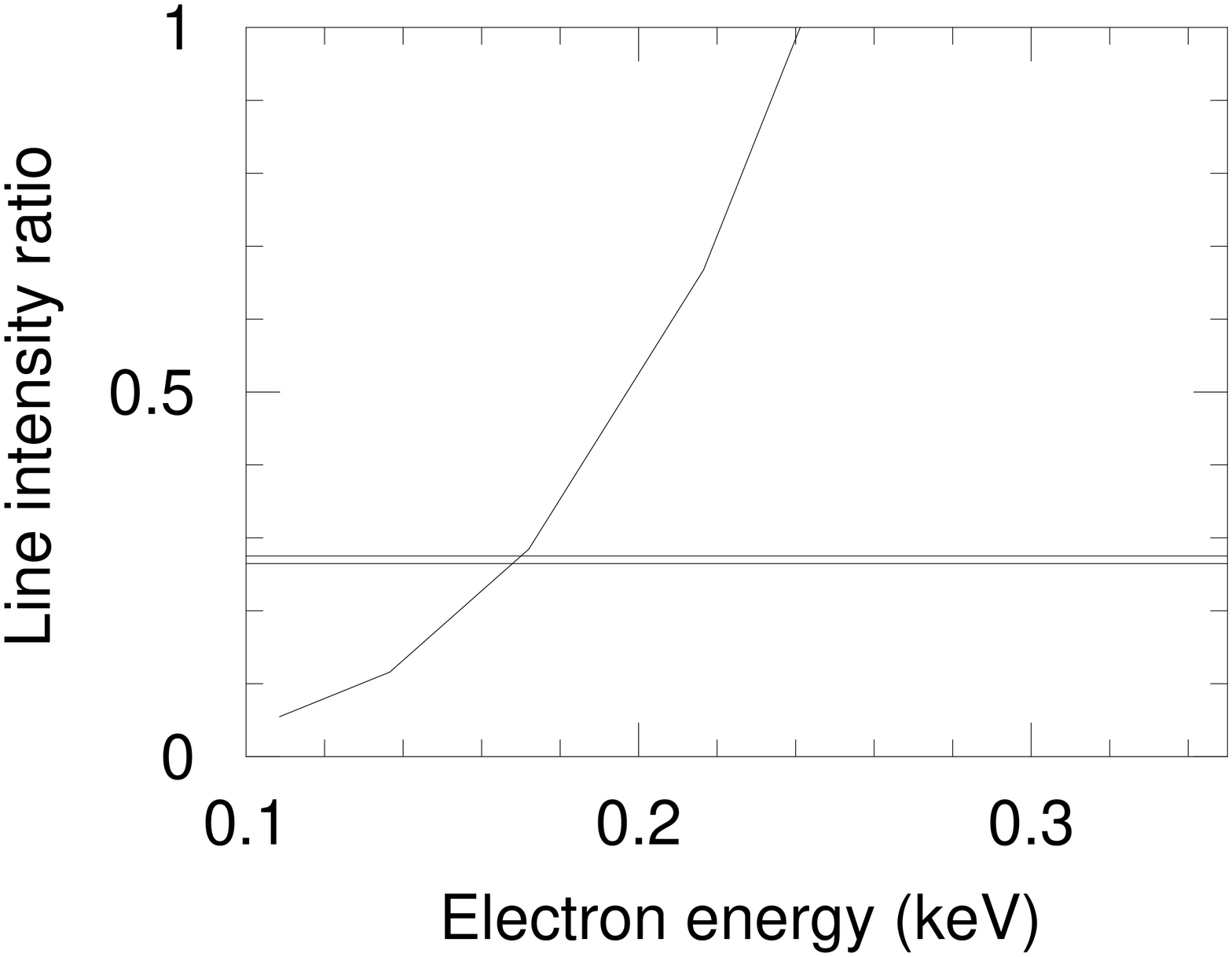}}
  \caption{Line intensity ratio of 0.63$-$0.67\,keV band to
  0.54$-$0.58\,keV band. Two horizontal lines show the upper and lower
  limits for the intensity ratio of 0.653\,eV line to 0.564\,eV line.}
  \label{fig:line_intensity_ratio:O}
 \end{figure}

 Figure~\ref{fig:spec} shows the extracted Suzaku XIS1 spectrum.
  Many emission lines are
 detected in the 0.2$-$1\,keV region.  To identify them,
 we fitted a parameterized spectral model consisting of 10 
 intrinsically narrow 
 Gaussian functions with an underlying thermal bremsstrahlung continuum
 all absorbed by a photoelectric absorption function.
 The resultant best-fit emission line parameters and line identifications
 are summarized in table~\ref{table:gaus_fit}.
 We clearly detect line emission from
 highly-ionized C, N, and O.  Based on this model fit, we
 calculate the line intensity ratios among different ionic species of N and O
 which enable us to estimate $kT_{\rm e}$ for the N and O emitting plasma.
 Assuming $kT_{\rm e} = 0.13$\,keV and collisional ionization
 equilibrium conditions,   the
 strongest lines around 0.426\,keV and  0.495\,keV are
 N$\;${\scriptsize\rmfamily{VI}}\,K$\alpha$ and
 N$\;${\scriptsize\rmfamily{VII}}\,Ly$\alpha$, respectively, although
 there are also many
 emission lines from other elements present near these energies (ATOMDB
 ver 1.3.1; Smith et al.~2001).
 For comparison to the observed spectral results we use  ATOMDB 
 to calculate the intensity of all emission
 lines from all elements in the two energy bands: 0.48$-$0.51\,keV and
 0.41$-$0.44\,keV, whose widths are intended to account, approximately, 
 for the XIS energy
 resolution.   Figure~\ref{fig:line_intensity_ratio:N} shows the 
 modeled line intensity ratio (0.48$-$0.51\,keV band divided by 
 0.41$-$0.44\,keV band) as a function of  $kT_{\rm e}$. 
 The two  horizontal
 lines show the upper and lower limits for the observed intensity ratio 
 between the 0.495\,keV and  0.426\,keV lines.   Based on this figure, we
 find that the electron temperature of the plasma 
 predominantly giving rise to the N emission is
 $kT_{\rm e} =  0.127\pm0.005$\,keV which is 
 consistent with that determined from the
 continuum emission (see table~\ref{table:gaus_fit}).
 We do a similar calculation for the
 energy bands around 0.564\,eV and 0.653\,eV.  
 The bands used for the model calculations here, 0.63$-$0.67\,keV and
 0.54$-$0.58\,keV, are dominated by  emission 
 from highly-ionized O and thus we only need to consider the
 O emission lines in ATOMDB.
 Figure~\ref{fig:line_intensity_ratio:O}
 shows the modeled line intensity ratio. The electron temperature of the 
 O-emitting
 plasma is $kT_{\rm e}=0.166\pm0.001$\,keV which is slightly higher than 
  that of the 
 continuum emission and N-emitting plasma.  Given the 
  significant dependence of the N and O line
  emissivities on temperature in this low  $kT_{\rm e}$ regime, the 
  results just
 presented suggest that there are at least 
 two plasma components in this region.

\begin{figure}[htbp]
  \centering
  \rotatebox{-90}{\includegraphics[clip,scale=0.5]{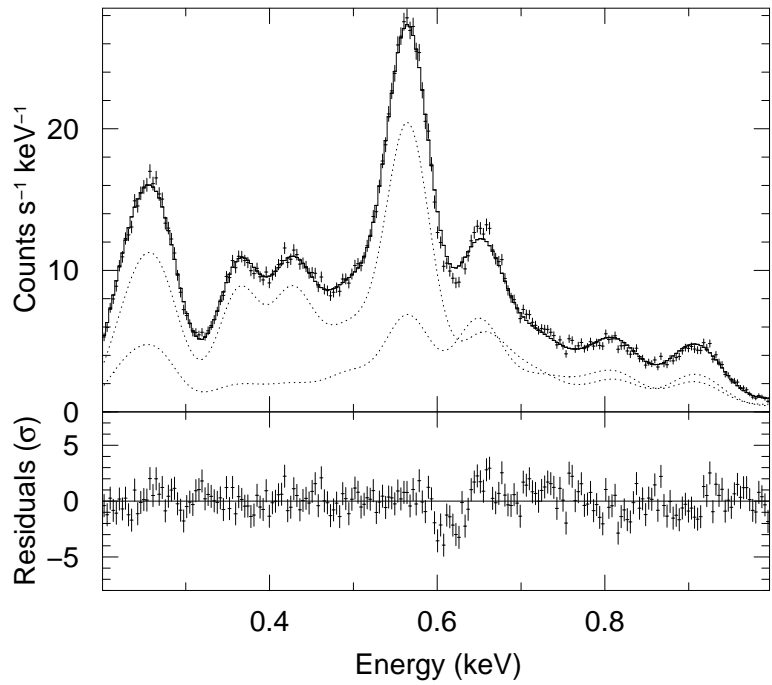}}
  \caption{Same as figure~\ref{fig:spec} but with the two component vnei model
  having different $kT_{\rm e}$ and different log($n_{\rm e}t$).}
  \label{fig:vnei_vnei}
 \end{figure}

 We are thus led to investigate a two-component  non-equilibrium 
  ionization (NEI)
 model. We used {\tt Xspec} ver 12.3 and the {\tt vnei} NEI
 model  (Borkowski et al.~2001) with the version number for the NEI
 models ({\tt NEIVERS}) of 2.0. 
 The free parameters are $kT_{\rm e}$,
 ionization timescale (log($n_{\rm e}t$)
 where $n_{\rm e}$ is the electron density 
 and $t$ is the time since the plasma was shock heated),
 emission measure, the absorbing column density, $N_{\rm H}$, and the
 abundances of C, N, O, Ne, and Fe relative to their solar
 values (Anders \& Grevesse 1989). The abundances of Mg, Si, S, Ar, and Ca
 were linked to that of Ne, while the Ni abundance was linked
 to Fe. $N_{\rm H}$ and abundances of 
 the heavy elements were kept
 the same for both plasma components. 
  The data are well
 reproduced by the {\tt vnei} model as can been seen in 
 figure~\ref{fig:vnei_vnei}, which plots
 the best-fit model in comparison to the spectral data.
 The best-fit  parameter values   are summarized in
 table~\ref{table:vnei}. 
  In paper I (see figure~5) there was extra scatter in the residuals plot 
  associated
 with the emission lines, which has  been greatly reduced
 in the present study thanks to our use of newer response
 matrix files that take into account the degradation of the XIS energy 
 resolution.

 The abundances of the heavy elements (with the exception of Si)
 are systematically larger than those presented
  in paper I. Updated calibration and response matrices files likely
 accounts for  some of this discrepancy.  The details of the model
  used for the fits differ as well.  Here we use two NEI models with 
independent temperatures and ionization timescales, whereas before
the  ionization timescales were the same for the two plasma components. 
 Another major difference is that here 
 the abundances of Ne,  Mg, Si, S, Ar, and Ca vary as a group with
 their relative abundances fixed at the solar values.  In
 paper I the abundances of Ne, Mg, and Si varied freely, while Ar and
 Ca were fixed at their full solar values.
 It is important to note that the abundances we derive here for nearly
 all the heavy elements are $\sim$0.23, which means that their relative
 abundances are consistent with the solar values.  This result strongly
 suggests that the soft X-ray emitting plasma is predominantly of
 interstellar origin.  However, O  stands out as being relatively
 depleted by a factor of two even though the O lines are among the
 strongest  ones in this spectral range,  as shown in 
 table~\ref{table:gaus_fit}.

 \section{Discussion and Conclusion}

 We analyzed the very soft X-ray emission from the northeastern region
 of the Cygnus Loop with the Suzaku Observatory.
 The X-ray emitting plasma requires at least two spectral
 components; a high $kT_{\rm e}$ component with small ionization timescale
 and a low  $kT_{\rm e}$ component with large ionization timescale.

 Usually, X-ray emitting plasmas of SNRs are assumed to be optically thin
 because the typical density is quite low.    
 Since the region we analyzed is
 very bright with a potentially large line-of-sight path length, however, this 
 assumption may not be valid for some resonance lines.
 The line-center cross section of resonance scattering can be expressed as 
 \begin{equation}%\scriptsize
  \sigma = {\sqrt{\pi} e^2 \over m c}{f \over \nu}\left({v \over c}\right)^{-1} \simeq 1.86 \times 10^{-9}\,{f \over E}\,v^{-1} \quad\mbox{cm}^2\ , 
  %=  1.85\times 10^{-9} f\,E^{-1} (\Delta v)^{-1}\,\,\,\, {\rm cm}^2\,\,,
 \end{equation}
 where $f$ and $\nu$ are the oscillator strength and frequency,
 respectively, of the line transition concerned, $v$ is the
 root-mean-square kinetic
 velocity of the ion, $m$ is the electron mass and other quantities have
 their usual meanings.  In the last expression, $E$ is the line energy
 in keV. The kinetic velocity is composed of thermal and turbulent
 motions, as 
 \begin{equation}%\scriptsize
 v^2 = \left({2kT_{\rm i} \over m_{\rm i}}\right)^2 + \xi^2\ ,
%  \Delta v =  \sqrt{\frac{2kT_{\rm i}}{m_{\rm i}}}\,\,\,\,\,
%   {\rm cm\, s}^{-1} \,\,,
 \end{equation}
 where $kT_{\rm i}$ is the ion temperature in keV, $m_{\rm i}$ is the ion
 mass, and $\xi$ represents the  root-mean-square turbulent velocity due to
 motions other than thermal one.  In the following analysis, we assume
 that the second term (turbulent motion) is negligibly small compared to
 the first one (thermal motion) in equation (2); the validity is
 discussed later.
 
  The line-center optical depth for resonance scattering is given by
 \begin{equation}%\scriptsize
 \tau = n_z \sigma L = \left({n_z \over n_Z}\right) \left({n_Z \over n_{\rm H}}\right) \left({n_{\rm H} \over n_{\rm e}}\right)\,n_{\rm e} \sigma L \ ,
%  \tau = 
%   \sigma
%   A
%   (Z/Z_{\rm H})
%   n_{\rm z}
%   L
%   n_{\rm e}\,,
 \end{equation}
where $L$ is the path length through the plasma, $n_z$ the ion density, $n_Z$ the element density, $n_{\rm H}$ the hydrogen density, and therefore $n_z/n_Z$ represents the ionic fraction, $n_Z/n_{\rm H}$ the elemental abundance relative to hydrogen.
% \noindent
% where $A$ is  the solar
% abundance of element, $Z/Z_{\rm H}$ is the elemental abundance
% relative to H, $n_{\rm z}$ is the ionic fraction, $L$ is the path length
% through the plasma, and $n_{\rm e}$ is the electron density. 
 We assume $L = 2.5$\,pc and the emission volume to be 2.5$\times$0.93$\times$2.5\,pc$^3$. 
 The electron density is obtained from the emission measure to be 1.25 and 1.35\,cm$^{-3}$ for the high and low $kT_{\rm e}$ components, respectively, and $n_{\rm e}/n_{\rm H}$ is taken to be 1.2.  For the ion temperature, we assume $kT_{\rm i} = kT_{\rm e}$ = 0.236\,keV,  which is the density-weighted mean value from table~\ref{table:vnei}.
 In the calculation of $\sigma$ for the He-like K$\alpha$ lines, we summed 
 the oscillator strengths for the three main lines of the He-like
 triplet, namely the forbidden ($z$), intercombination ($x+y$), and resonance
  ($w$) lines.
 The ionic fractions of
 O$\;${\scriptsize\rmfamily{VII}}\,K$\alpha$ and
 O$\;${\scriptsize\rmfamily{VIII}}\,K$\alpha$ were calculated by taking
 into account the NEI condition for each component separately (using Masai 
1984).  The density-weighted overall ionic fractions of
 O$\;${\scriptsize\rmfamily{VII}}\,K$\alpha$ and
 O$\;${\scriptsize\rmfamily{VIII}}\,K$\alpha$ were then determined to be 
 0.51 and 0.33, respectively.  The
 inferred values of $\tau$ for
 O$\;${\scriptsize\rmfamily{VII}}\,K$\alpha$ and
 O$\;${\scriptsize\rmfamily{VIII}}\,K$\alpha$ turn out to be 0.54 and 0.16.
 Table~\ref{table:optical_depth} summarizes the resonance-line-scattering
 cross sections and optical depths for the K emission lines detected in
 the Suzaku data and 
 shown in table~\ref{table:gaus_fit}.  The optical depth of
 O$\;${\scriptsize\rmfamily{VII}}\,K$\alpha$ is the largest whereas
 those of the other emission lines are not negligibly small.

 In order to model the effect of self-absorption in a simple way, we
 employed the so-called ``escape-factor'' method (Irons 1979).  We assume
 the global (source- and direction-average) Doppler-profile escape
 factor for plane-parallel ``slab'' geometry (Bhatia \& Kastner 2000)
 which is a reasonable assumption for the shell-brightened appearance of
 the Cygnus Loop.  Escape factor values for the relevant emission lines
 are summarized in table~\ref{table:optical_depth}.
 All values are less than 0.87, indicating that our line intensities have been
 underestimated by a factor of 13\% or more.
 Optical depth effects, therefore, play a significant  role in the X-ray
 emission from the Cygnus Loop with the O abundance, in particular,
 being underestimated by a factor of 20--40\%.  However, optical depth
 effects alone cannot account for the entire abundance discrepancy of a
 factor of two that we observe.  It should be noted as well that the
 escape factor is different for
 O$\;${\scriptsize\rmfamily{VII}}\,K$\alpha$ and
 O$\;${\scriptsize\rmfamily{VIII}}\,K$\alpha$, implying that the {\it true}
 line intensity ratio differs from the observed value.
 Correcting the line intensity ratio of 
 O$\;${\scriptsize\rmfamily{VIII}}\,K$\alpha$ to
 O$\;${\scriptsize\rmfamily{VII}}\,K$\alpha$ for this effect results
 in a revised ratio of  0.21, which implies 
 $kT_{\rm e}=0.15$\,keV based on the APEC model of equilibrium
 ionization.  This optical
 depth effect not only reduces  derived elemental abundances but it
 also biases fitted $kT_{\rm e}$ measurements from their {\it true} 
 values.  Furthermore, varying the 
 elemental abundance affects the calculation of the
 optical depth as shown in equation (3).
 Future work will be needed to develop more sophisticated emission
 codes that take optical depth effects into account.

 We have made a few assumptions that tend to enhance the effects of
 resonance line scattering, including
 (a) temperature equipartition between electrons and ions,
 (b) no bulk motions, and
 (c) no turbulence.
 The temperature equipartition is a reasonable assumption for the region observed, as argued by Ghavamian et al. (2001) based on the optical
 spectra as well as by Miyata
 \& Tsunemi (1999) on X-rays.
 Bulk motions may not be such a big problem for the Cygnus Loop
 due to its relatively low shock speed (few hundred km\,s$^{-1}$) and
 its large angular size on the sky.
 The region observed is right at
 the edge of the remnant's rim and therefore can be approximated
 as a slab moving uniformly
 across the sky.  This would not be the case if there was
 significant curvature of the remnant shell in the line of sight,
 or if we were viewing through the center of the remnant where the
 back part of the shell would be moving away from us while the front
 part was moving toward us.  So the proximity of the Cygnus Loop to
 the Earth again favors resonance line scattering.
%% The observing region
%% is right at the edge and is very nearly all moving across the sky 
%% as a slab.  So the back part of the slab is moving in nearly the
%% same direction and at the same speed as the front part of the
%% slab.  This would not be the case if there was significant curvature
%% of the remnant shell in the line of sight.  So the proximity of
%% the Cygnus Loop to Earth again favors resonance line scattering.

 As for (c), very little information is available.  One location where
 turbulence likely occurs  is at the
 contact discontinuity of the ejecta in young SNRs.  The instabilities
 that operate here drive turbulent eddies and vortices that imprint a
 random pattern of velocity flows onto the overall radial expansion of
 the ejecta.  Any such turbulence has an effect of increasing the
 root-mean-square velocity broadening $v$ resulting in a reduction
 of the scattering cross section.

 The forward shock in the adiabatic phase of remnant evolution is
 hydrodynamically stable and will not suffer this effect.  But radiative
 cooling at the forward shock causes hydrodynamical instabilities
 (Blondin et al. 1998).  Ultimately the amount of line
 broadening due to turbulence of this type will be some fraction of the
 shock velocity, so that younger SNRs will tend to have a
 higher line broadening and hence lower cross-section for resonance line
 scattering.  If the part of the Cygnus Loop we observed is still
 in the Sedov phase of evolution, the conditions for minimal turbulence
 are met.  Therefore, according to all three items discussed above,
 the Cygnus Loop is favored to have a significant effect due to resonance
 line scattering.

%%%%%%%%%%%%%%%%%%%%%%%%%%%%%%%%%%%%%%%%%%%%%%%%%%%%%%%%%%%%%%%%%%%%%%

 Since the Cygnus Loop appears to be a rim-brightened shell in the X-ray 
 band,
 optical depth effects may only play an important role at the rim where
 the plasma path lengths are largest.   The abundance of O relative to other 
 heavy elements has been show to decrease at the shell region of the Cygnus 
 Loop (e.g. Miyata \& Tsunemi 1999).  Likewise, Leahy (2004)
 analyzed Chandra data of the bright southwestern region and found
 that the abundances of O group elements (C, N, and O) is roughly half
 those of the 
 Ne group (Ne, Na, Mg, Al, Si, S, and Ar) and Fe group (Ca, Fe, and Ni). 
 The depleted O abundance relative to other elements is
 consistent with our results and suggests that
 optical depth effects might be important at the southwestern region too.
%% It should be noted that the abundance of O group derived at 
%% the relatively dim region also shows the depleted values whereas the
%% statistical errors are large.  This fact suggests that O
%% might be systematically depleted at the shell region with any other
%% unknown reason.
 It should be noted that Miyata \& Tsunemi (1999) and Leahy
 (2004) assumed the C and N abundances relative to O to be solar 
 because they had neither high enough spectral resolution nor sufficient
 effective area for detecting C and N lines.  The Suzaku XIS camera 
 enables us to
 determine the abundances of C, N, and O independently for diffuse
 X-ray sources so future additional
 observations with Suzaku are essential for abundance measurements.
 Katsuda et al.~(2007) analyzed four Suzaku pointings of the
 northeastern region of the Cygnus Loop
 and showed that O is relatively depleted by a factor of two
 compared  with other heavy elements.  This result also
 supports our view for a significant optical depth effect.
 The Suzaku observatory has so far performed 8 mapping observations from
 the northeastern toward the southwestern region of the Cygnus Loop.
 Detailed analysis
 of the radial distribution of intensities of
 O$\;${\scriptsize\rmfamily{VII}}\,K$\alpha$  and
 O$\;${\scriptsize\rmfamily{VIII}}\,K$\alpha$ as well as their intensity
 ratios may help to clarify optical depth effects in the SNR.  
%
%% However, the plasma structure at the inner region is not uniform along
%% the line of sight.  Such non-uniformity also affects the line intensity
%% rations.  It might be difficult to clarify the optical depth effect by
%% measuring the radial variation of the line intensity ratio.
 However, for this to work it will be necessary to separate out the
 effects of any intrinsic radial variation in the plasma temperature,
 which can also produce variations in the observed line intensity ratio.

 In clusters of galaxies, an anomalous ratio of Fe K$\beta$ to K$\alpha$
 emission lines has been observed, which is offered as evidence for
 resonance line scattering in the relatively dense central cluster plasma
(e.g., Tawara 1995).  Unfortunately, due to their NEI condition,
 the intensity ratio of the  K$\beta$ to K$\alpha$
 emission line complexes is not a good indicator of 
 resonance line scattering in SNRs.
 For definitive answers,  we need to resolve individual K-shell lines of 
 He-like ions  as well as the  Ly$\alpha$ and Ly$\beta$ lines  for a
 range of elemental species as  a function of position across the
 extent of diffuse thermal X-ray sources.  Hopefully in the near future
 X-ray microcalorimeters onboard $XEUS$, Con-X, and/or $NeXT$  will provide
 this capability and allow X-ray astronomers to use resonance scattering
 and other techniques to gain a deeper insight into the nature of SNRs.

 \vspace{1cm}

 We are grateful to all the other members of the Suzaku team.
 EM is supported by  Grant-in-Aid for Specially Promoted
 Research (16002004) and 
 the 21st Century COE Program, \lq{\it Towards a new basic
 science: depth and synthesis}\rq. JPH acknowledges support from
 NASA grant NNG05GP87G.
 This research has made use of data
 obtained through the High Energy Astrophysics Science Archive
 Research Center Online Service, provided by the NASA/Goddard Space
 Flight Center.

 \clearpage

 \begin{table}[htbp]
  \caption{Best-fit parameters of the Gaussian fitting.}
  \label{table:gaus_fit}
  \centering
   \begin{tabular}{lll} \hline \hline
    Line energy (eV) & Intensity (photons s$^{-1}$ cm$^{-2}$) &
    Line identification \\ \hline
    358$\pm$5 & (3.0$\pm$0.1)$\times$10$^{-2}$ &
    C$\;${\scriptsize\rmfamily{VI}}\,K$\alpha$ \\
    426$\pm$5 & (1.1$\pm$0.1)$\times$10$^{-2}$ &
    N$\;${\scriptsize\rmfamily{VI}}\,K$\alpha$\\
    495$\pm$5 & (3.9$\pm$0.2)$\times$10$^{-3}$ &
    N$\;${\scriptsize\rmfamily{VII}}\,Ly$\alpha$, N$\;${\scriptsize\rmfamily{VI}}\,K$\beta$\\
    564$\pm$5 & (3.20$\pm$0.05)$\times$10$^{-2}$ &
    O$\;${\scriptsize\rmfamily{VII}}\,K$\alpha$\\
    653$\pm$5 & (8.7$\pm$0.1)$\times$10$^{-3}$ &
    O$\;${\scriptsize\rmfamily{VIII}}\,Ly$\alpha$\\
    722$\pm$5 & (2.8$\pm$0.1)$\times$10$^{-3}$ &
    Fe$\;${\scriptsize\rmfamily{XVII}}\\
    792$\pm$6 & (1.9$\pm$0.1)$\times$10$^{-3}$ &
    Fe$\;${\scriptsize\rmfamily{XVII}}, Fe$\;${\scriptsize\rmfamily{XVIII}}\\
    837$\pm$7 & (1.2$\pm$0.1)$\times$10$^{-3}$ &
    Fe$\;${\scriptsize\rmfamily{XVII}}, O$\;${\scriptsize\rmfamily{VIII}}\,Ly$\beta$\\
    910$\pm$5 & (2.01$\pm$0.06)$\times$10$^{-3}$ &
    Ne$\;${\scriptsize\rmfamily{IX}}\,K$\alpha$\\
    989$\pm$9 & (2.6$\pm$0.7)$\times$10$^{-4}$ &
    Fe$\;${\scriptsize\rmfamily{XVII}}, Ni$\;${\scriptsize\rmfamily{XIX}}\\
    \hline
    Thermal bremsstrahlung & $kT_{\rm e}$ (keV) & 0.13$\pm$0.01 \\
    Absorption & $N_{\rm H}$ ($\times 10^{20}$ cm$^{-2}$) & 1.9$\pm$0.2 \\
    $\chi^2$ (d.o.f.)$^*$ & 205 (195) & \\
    \hline
   \end{tabular}
  \begin{flushleft}
  \noindent
  Errors quoted are 90\% confidence level.\\
  \noindent
  $^*$ d.o.f. means degree of freedoms.
  \end{flushleft}
 \end{table}

 \begin{table}[htbp]
  \caption{Best-fit parameters of the {\tt vnei} model fitting.}
  \label{table:vnei}
  \centering
   \begin{tabular}{ll} \hline \hline
    \multicolumn{2}{l}{High $kT_{\rm e}$ component} \\
    \hline
    $kT_{\rm e}$ (keV) & 0.248$^{+0.006}_{-0.001}$ \\
    log($n_{\rm e}t$) &  10.6$^{+0.02}_{-0.03}$ \\
    Emission measure$^*$ (cm$^{-6}$pc$^3$)& 7.6$\pm$0.1\\
    \hline
    \multicolumn{2}{l}{Low $kT_{\rm e}$ component}  \\
    \hline
    $kT_{\rm e}$ (keV) & 0.224$\pm$0.001 \\
    log($n_{\rm e}t$) & 11.98$\pm$0.06  \\
    Emission measure$^*$ (cm$^{-6}$pc$^3$) & 8.8$^{+1.9}_{-0.2}$\\
    \hline
    Element & Abundance \\ \hline
    C & 0.22$\pm$0.01 \\
    N & 0.24$\pm$0.01 \\
    O & 0.13$\pm$0.01 \\
    Ne,Mg,Si,S,Ar,Ca & 0.23$\pm$0.01\\
    Fe,Ni & 0.24$\pm$0.01 \\
    \hline
    $N_{\rm H}$ ($\times 10^{20}$ cm$^{-2}$) & $<$\,2 \\
    $\chi^2$ (d.o.f.) & 314 (204) \\
    \hline
   \end{tabular}
  \begin{flushleft}
  \noindent
   Errors quoted are 90\% confidence level.\\
   $^*$ Emission measure is defined as $\int n_{\rm e} n_{\rm H} V$
   where $n_{\rm H}$ is the Hydrogen density and $V$ is the emission volume.
  \end{flushleft}
 \end{table}

 \begin{table}[htbp]
  \caption{Cross section of resonance line scattering, optical depth, and
  escape factor for K emission lines detected.}
  \label{table:optical_depth}
  \centering
   \begin{tabular}{lccc} \hline \hline
    Emission line identified & Cross section (10$^{-16}$cm$^2$)&
    Optical depth & Escape factor  \\ \hline
    C$\;${\scriptsize\rmfamily{VI}}\,K$\alpha$ & 3.4 & 0.17 & 0.82 \\
    N$\;${\scriptsize\rmfamily{VI}}\,K$\alpha$ & 6.0 & 0.11 & 0.87 \\
    O$\;${\scriptsize\rmfamily{VII}}\,K$\alpha$ & 4.8 & 0.54 & 0.63 \\
    O$\;${\scriptsize\rmfamily{VIII}}\,Ly$\alpha$ & 2.2 & 0.16 & 0.83 \\
    Ne$\;${\scriptsize\rmfamily{IX}}\,K$\alpha$ & 3.3 & 0.17 & 0.82 \\
    \hline
   \end{tabular}
 \end{table}

 \clearpage

 \clearpage

 \clearpage

 \clearpage

 \clearpage

\end{document}